\documentclass[manuscript,review=false,anonymous=false]{acmart} 

\AtBeginDocument{%
  \providecommand\BibTeX{{%
    \normalfont B\kern-0.5em{\scshape i\kern-0.25em b}\kern-0.8em\TeX}}}

\setcopyright{acmcopyright}
\copyrightyear{2018}
\acmYear{2018}
\acmDOI{10.1145/1122445.1122456}

\usepackage{booktabs} 
\usepackage{graphicx} 
\usepackage{url}

\usepackage{caption}
\usepackage[all]{nowidow}
\usepackage{wrapfig}
\usepackage{array}
\usepackage{arydshln}
\usepackage{tabularx}
\usepackage{multirow}
\usepackage{arydshln}
\newcolumntype{L}[1]{>{\raggedright\let\newline\\\arraybackslash\hspace{0pt}}m{#1}}
\newcolumntype{C}[1]{>{\centering\let\newline\\\arraybackslash\hspace{0pt}}m{#1}}
\newcolumntype{R}[1]{>{\raggedleft\let\newline\\\arraybackslash\hspace{0pt}}m{#1}}

\usepackage{wrapfig,lipsum,booktabs} 

\def\authnotes{1}
\newcounter{notectr}[section]
\newcommand{\thenote}{\thesubsection.\arabic{notectr}\refstepcounter{notectr}}

\newcommand{\note}[2]{$\ll$#1~\thenote: #2$\gg$}
\newcommand{\cnote}[1]{\ifnum\authnotes=1 \textcolor{blue}{\note{Comment:}{#1}}\fi}

\acmYear{2026}
\copyrightyear{2026}
\acmConference[CHI '26]{CHI}{December 9--11, 2026}{USA}
\acmBooktitle{CHI'26, December 9--11, 2024, USA}
\acmDOI{10.1145/3700794.3XXXXXX}
\acmISBN{979-8-4007-1041-4/24/12}

\begin{document}



\title[Bangladesh AI Readiness]{Bangladesh's AI Readiness: Perspectives from the Academia, Industry, and Government}

\author{Sharifa Sultana}
\affiliation{%
  \institution{University of Illinois Urbana-Champaign}
  \country{USA}}
\email{sharifas@illinois.edu}

\author{Rupali Samad}
\affiliation{%
  \institution{University of Dhaka}
  \country{Bangladesh}}
\email{bsse1208@iit.du.ac.bd}

\author{Mehzabin Haque}
\affiliation{%
  \institution{University of Dhaka}
  \country{Bangladesh}}
\email{bsse1233@iit.du.ac.bd}

\author{Zinnat Sultana}
\affiliation{%
  \institution{S.M.R. Law College, Jessore}
  \country{Bangladesh}}
\email{zinnat1409@gmail.com}

\author{Zulkarin Jahangir}
\affiliation{%
  \institution{North South University}
  \country{Bangladesh}}
\email{zulkarin@gmail.com}

\author{B M Mainul Hossain}
\affiliation{%
  \institution{University of Dhaka}
  \country{Bangladesh}}
\email{}

\author{Rashed Mujib Noman}
\affiliation{%
  \institution{Augmedix Bangladesh}
  \country{Bangladesh}}
\email{r.m.noman@gmail.com}

\author{Syed Ishtiaque Ahmed}
\affiliation{%
  \institution{University of Toronto}
  \country{Canada}}
\email{ishtiaque@cs.toronto.edu}

\renewcommand{\shortauthors}{Sultana et al.}

\begin{abstract}
Artificial Intelligence (AI) readiness in the Global South extends beyond infrastructure to include curriculum design, workforce development, and cross-sector collaboration. Bangladesh, ranked 82nd in the 2023 Oxford Insights AI Readiness Index, exhibits significant deficits in technology capacity and research ecosystems, despite strong governmental visions. While HCI and ICTD research have explored digital inclusion and responsible AI, little empirical work examines how educational, industrial, and policy domains intersect to shape readiness. We present a multi-method qualitative study of AI readiness in Bangladesh, combining institutional analyses, 59 stakeholder interviews, and curriculum benchmarking against global exemplars. Findings reveal outdated curricula, limited faculty upskilling, inadequate computing resources, entrenched gender disparities, and the near-total absence of AI ethics instruction. We contribute empirical mapping of current practices, identification of structural and cultural barriers, and actionable pathways for embedding human-centered, inclusive, and responsible AI practices into national agendas, advancing equitable innovation in emerging AI ecosystems.

\end{abstract}







\settopmatter{printfolios=true}

\maketitle

\section{Introduction}

Artificial Intelligence (AI) is increasingly recognized as a transformative force across education, industry, and governance. For countries in the Global South, including Bangladesh, AI adoption is tied not only to technological advancement but also to broader socio-economic development goals. Yet, AI readiness extends beyond hardware and software capabilities—it encompasses the institutional capacity to integrate AI into curricula, the cultivation of an ethically aware workforce, and cross-sectoral collaboration that ensures AI solutions address local priorities. The 2023 Oxford Insights AI Readiness Index ranks Bangladesh 82nd globally, with notable strengths in governmental vision but persistent deficits in technology infrastructure and research ecosystems. This imbalance signals a pressing need to understand how academic, industrial, and policy actors can collaboratively close readiness gaps.

Within HCI, AI readiness in Global South contexts remains underexamined. While prior ICTD and AI ethics research has investigated infrastructure disparities, digital inclusion, and culturally responsive AI design, there is limited empirical work mapping the intersection of curriculum adequacy, faculty development, and cross-sector engagement in emerging AI ecosystems. Current HCI literature tends to focus either on individual user adoption or high-level governance frameworks, leaving a gap in understanding the everyday institutional and pedagogical practices that mediate AI readiness. In Bangladesh, where AI strategy documents and policy drafts reference global ethical frameworks like UNESCO’s Recommendation on the Ethics of AI, little is known about how such aspirations translate into actual educational and industrial capacity-building.

This paper addresses that gap through a multi-method qualitative study of AI readiness across academia, industry, and government in Bangladesh. Drawing on the aiEDU Readiness Framework and UNESCO AI Competency Standards, we conducted institutional analyses, semi-structured interviews with 59 stakeholders, two targeted surveys, and curriculum benchmarking against global exemplars. Our broader goal is to situate Bangladesh’s AI readiness within both local constraints and global discourses, identifying actionable levers for HCI researchers, educators, and policymakers. Specifically, we address three research questions: 

\begin{quote}
\textit{RQ1:} How do academic curricula and faculty development practices align with global AI trends and industry needs? \\
\textit{RQ2:} What structural, infrastructural, and cultural barriers shape cross-sectoral collaboration in AI education and deployment? \\
\textit{RQ3:} How are gender, ethics, and inclusivity addressed—or overlooked—in current AI readiness efforts?
\end{quote}

Our findings reveal systemic misalignments between academic instruction and industry expectations, driven by outdated curricula, limited exposure to emerging AI techniques, and scarce access to high-performance computing resources. Faculty and students lack consistent opportunities for training, mentorship, and interdisciplinary collaboration, while gender disparities remain entrenched across both academia and industry. Ethical considerations and Responsible AI principles are almost entirely absent from formal instruction, despite their prominence in policy rhetoric. These gaps result in a skills pipeline that is underprepared for both the technical and social complexities of contemporary AI work.

We offer four key contributions to HCI from this study. \textbf{First}, we provide empirical evidence by mapping current practices and growth agendas of Bangladeshi AI from academia and industry perspectives. \textbf{Second}, we highlight the critical absence of ethics and gender inclusivity in AI education, emphasizing the implications for equitable innovation. \textbf{Third}, we extend AI and ICTD scholarship by foregrounding the role of institutional culture and cross-sector inertia in shaping technological capacity-building. \textbf{Fourth}, we propose actionable pathways for integrating human-centered, socially responsible AI practices into national AI readiness agendas, bridging the gap between policy aspirations and lived educational realities.
\section{Related Work}

National AI readiness is commonly assessed through composite indices that emphasize state capacity, technology infrastructure, and data ecosystems. The Oxford Insights Government AI Readiness Index (2023) situates Bangladesh in the lower-middle tier globally, underscoring gaps in the technology pillar despite comparatively stronger government vision and policy intent~\cite{oxfordinsights2023}. At the same time, global normative frameworks are maturing: UNESCO’s Recommendation on the Ethics of AI (2021) provides high-level principles for human rights–centred AI governance~\cite{unesco2021ethics}, while the EU AI Act (adopted 2024) introduces binding, risk-based obligations for AI systems—together shaping a policy environment in which readiness increasingly entails ethical, legal, and institutional capacity, not only hardware and talent~\cite{euaiact2024}.

\subsection{AI Readiness and HCI}

HCI and adjacent ICTD research have called attention to the socio-technical substrates of “readiness,” documenting how data labor, institutional incentives, and power asymmetries shape AI practice in emerging contexts. Sambasivan et al. show that data cascades—systematic, compounding issues arising from data collection and stewardship—undermine deployed AI systems and are frequently overlooked relative to model-building work~\cite{sambasivan2021datawork}. Relatedly, work on re-imagining algorithmic fairness in India foregrounds local values, governance capacity, and contested notions of harm, arguing that fairness cannot be parachuted from the Global North without re-specifying problem framings and accountability~\cite{sambasivan2021fairnessIndia}. These strands collectively reframe readiness as a situated organizational and cultural capacity—aligning with our focus on curricula, faculty development, and cross-sector engagement rather than solely national strategy documents. 

Within education, a rapid literature on responsible AI converges on two patterns: (1) universities worldwide are moving from ad-hoc responses to generative AI toward institution-wide guidance and (2) most programs still lack systematic ethics integration across the curriculum. Reviews in AIED and higher-education studies chart both the pedagogical opportunities and the governance gaps around transparency, accountability, and assessment integrity~\cite{fu2024aiedu,dabis2024hepolicy}. Complementing this, UNESCO’s AI Competency Framework for Students emphasizes human-centred mindsets and ethics as core outcomes, and aiEDU’s readiness frameworks translate such competencies into implementation rubrics for districts and classrooms~\cite{unescoAICF2025, aiedu2025}. Our curriculum benchmarking and interview findings situate Bangladesh within these global trends while surfacing Global-South-specific constraints (e.g., GPU scarcity, informal mentorship networks).

\subsection{AI Readiness: Global Perspective}

Contemporary “AI readiness” extends beyond innovation capacity to encompass rights-preserving governance and institutional capability. Benchmarks like the Oxford Insights Government AI Readiness Index summarize national capacity across government, technology, and data pillars—in which Bangladesh currently lags on the latter two despite stronger policy intent \cite{oxfordinsights2023}. In parallel, global normative baselines have consolidated around UNESCO’s \emph{Recommendation on the Ethics of AI} (2021), a human-rights-centred instrument adopted by all Member States \cite{unesco2021ethics}, and the European Union’s AI Act (adopted May 2024; in force since August 1, 2024), which establishes a risk-based regulatory architecture with tiered obligations for providers and deployers \cite{euaiact2024}. Together, these instruments position readiness as compliance and accountability capacity—not only compute, data, and talent.

The EU AI Act operationalizes this shift by prohibiting selected practices and imposing requirements on “high-risk” systems (e.g., quality management, data governance, logging, human oversight, post-market monitoring), alongside staged application timelines that run into 2025–2026. This effectively rewrites skill needs for education and industry (e.g., risk management, documentation, impact assessment) and challenges universities to integrate governance competencies into technical training \cite{euaiact2024}.

Education policy has begun to translate these expectations into institutional guidance. UNESCO’s \emph{Guidance for Generative AI in Education and Research} (2023) recommends immediate actions (policy and procurement rules, educator capacity-building, assessment integrity) and long-term planning to ensure a human-centred adoption of GenAI across schools and universities \cite{miaoHolmes2023}. These steps connect national ethical principles to everyday classroom and research practice, and to resourcing decisions \cite{miaoHolmes2023}.

Despite this progress, policy implementation remains sparse: a UNESCO global survey of over 450 institutions found that fewer than 10\% had formal institutional guidance on GenAI as of 2023—evidence of a large governance and capacity gap even in well-resourced systems \cite{unesco2023less10}. This gap frames readiness as an institutional change agenda (policy, professional development, assessment redesign), not merely a technology adoption question.

Finally, multilaterals now provide “policy infrastructure.” The OECD AI Policy Observatory curates national strategies and tracks implementation of the OECD AI Principles (2019; refreshed 2024), supporting cross-country learning and interoperability—useful for aligning Bangladesh’s efforts with emerging global norms \cite{oecd2023trackingai}.

\subsection{AI Readiness in the Global South}

HCI/ICTD research shows that readiness is situated: it is co-produced by data practices, institutional incentives, and power dynamics rather than by strategy documents alone. Sambasivan et al. document “data cascades”—compounding data issues that degrade AI systems in high-stakes domains, often overshadowed by model-centric work \cite{sambasivan2021datawork}. Complementary research on contextual fairness in India argues that porting Western fairness templates without re-specifying harms, accountability, and community power can be counter-productive \cite{sambasivan2021fairnessIndia}. Taken together, these accounts recast readiness as organizational capacity for data stewardship, participatory problem framing, and contested decision-making, not just technical skill acquisition \cite{sambasivan2021fairnessIndia, sambasivan2021datawork}. 

Related work on AI authority in India further shows how aspirational attitudes toward AI can legitimate automation even when evidence of capability is thin—raising the bar for transparency, documentation, and public oversight in emerging contexts \cite{kapania2022authority}. These findings reinforce that Global-South readiness must include cultural and institutional competencies (e.g., critical evaluation, contestability) alongside curriculum updates.

Bangladesh reflects these dynamics. On the one hand, it has operational security capacity (e.g., BGD e-Gov CIRT) and improving e-government indicators \cite{bgdcirt, egdi2024}. On the other, the Cyber Security Act (2023) has drawn human-rights concerns, and comprehensive data-protection reform remains under debate—indicating gaps in algorithmic accountability and citizen safeguards that are integral to responsible AI deployment \cite{bdcsa2023, bdpdp2025}. Readiness therefore depends as much on legal-institutional plumbing as on technical assets.

To bridge policy aspiration and institutional reality, countries are adopting readiness toolkits. UNESCO’s Readiness Assessment Methodology (RAM)—implemented in collaboration with Bangladesh’s ICT Division and a2i—assesses legal/regulatory, social, educational, and infrastructural dimensions and yields actionable recommendations for cross-sector coordination \cite{unescoethicsbd}. Such diagnostics can anchor curriculum reform, workforce development, and procurement/governance updates, even under resource constraints.

\section{Methods}
Our mixed-method research involves desktop research and interviews to investigate the state of AI readiness in Bangladesh, particularly across the academic, industrial, and governance sectors.

\subsection{Desktop Research}
We assessed institutional capacity, curriculum adequacy, faculty development, infrastructure availability, and stakeholder engagement of 35 prominent computer science and related programs in Bangladesh. The universities were selected based on their rankings, relevance, and contribution to AI research, and presence of AI-related topics and class projects. These institutions represent varying governance models (public vs. private), funding structures, and levels of access to infrastructure—conditions that are emblematic of broader educational asymmetries in the Global South. Furthermore, we explored how collaboration across academia, industry, and government currently functions. In this regard, we collected and scrutinized their syllabus of courses related to AI; events, newsletters, and announcements put on their websites; and research profiles of their faculty members and researchers on their personal websites and Google Scholar. We looked into faculty members' citation counts, focusing on those in top AI conferences (e.g., NeurIPS, ICML, CVPR, AAAI, AIES, and ACL and NLP venues) and evaluated the number of faculty members actively publishing in these high-impact venues. We also took note of the gender ratio of faculty members teaching AI-related courses and evaluated the gender composition of AI industry teams across the universities.

\subsection{One-on-One Interview}
We conducted one-on-one semi-structured interviews with 59 participants to understand their perceptions of AI education quality, institutional readiness, and cross-sector engagement. The participant pool included students (n=25) and educators (n=9) from the five prominent computer science and related programs that teach AI in four Bangladeshi universities. They are the Computer Science and Engineering program at Bangladesh University of Engineering and Technology (CSE at BUET, henceforth), the Computer Science and Engineering program at the University of Dhaka (CSE at DU, henceforth), the Institute of Information Technology at the University of Dhaka (IIT at DU, henceforth), the Computer Science and Engineering program at North South University (CSE at NSU, henceforth), and Computer Science and Engineering program at BRAC University (CSE at BRACU, henceforth). We also interviewed professionals (n=25) currently working in the local Bangladeshi computing and related industry. Student and faculty participants were recruited via university mailing lists, social media, and peer referrals. Industry experts were identified through alumni networks and prior collaborative projects with academia. We employed purposive sampling strategies and deliberately chose participants or cases based on their specific characteristics, knowledge, or experiences that are relevant to the study \cite{campbell2020purposive}. Note that this recruitment strategy was not aimed at statistical generalizability, but rather at capturing diverse, situated perspectives on AI education and collaboration practices in the Bangladeshi context. Purposive sampling is widely used in qualitative research and studies with small sample sizes, where the goal is not to generalize findings to a larger population but to gain an in-depth understanding of a specific phenomenon or group, which aligns with the sentiments and goals of our study.

Interview protocols were informed by two internationally recognized frameworks: the \textit{aiEDU Readiness Framework} \cite{ai_readiness_framework_2024}, which emphasizes curriculum, ethics, and infrastructure, and the \textit{UNESCO AI Competency Framework} for teachers and students, which foregrounds interdisciplinary and socially responsible AI education \cite{Miao_Cukurova_2024, Miao_Shiohira_Lao_2024}. The interviews with the students explored the accessibility and relevance of AI coursework, perceived gaps in hands-on experience, mentorship availability, and awareness of ethical implications. The interviews with the faculty members focused on personal pedagogical growth, skill development training and knowledge sharing through workshops and conferences across the nation and cross-border, curriculum design challenges, access to computing resources, and institutional support for AI pedagogy. Interviews with the industry personnel probed expectations of AI graduates, experiences of academic collaboration, and barriers to offering internships or co-designed curricula. Each interview lasted 45–60 minutes and was conducted in Bangla, as all of the participants preferred Bangla. We recorded the interview sessions with the permission of the participants, and later transcribed and anonymized them for further analysis. 

\begin{table}[!t]
\centering
\caption{Participant Demographics by Institution and Role}
\begin{tabular}{|l|l|c|c|c|}
\hline
\textbf{Participant Type} & \textbf{Institution/Org} & \textbf{Gender (M/F)} \\
\hline
\multirow{5}{*}{Undergraduate Students} 
  & BUET       & 5 (M:4; F:1)  \\
  & DU CSE     & 5 (M:3; F:2) \\
  & IIT DU     & 5 (M:3; F:2) \\
  & BRACU      & 5 (M:3; F:2) \\
  & NSU        & 5 (M:3; F:2) \\
\hline
\multirow{5}{*}{Faculty Members} 
  & BUET       & 2 (M:2; F:0) \\
  & DU CSE     & 2 (M:2; F:0) \\
  & IIT DU     & 2 (M:2; F:0) \\
  & BRACU      & 2 (M:2; F:0) \\
  & NSU        & 1 (M:1; F:0) \\
\hline
\multirow{5}{*}{Industry Professionals} 
  & Samsung R\&D BD     & 5 (M:4; F:1) \\
  & Streams Tech Ltd.   & 5 (M:5; F:0) \\
  & Cefalo              & 5 (M:4; F:1) \\
  & Factorize           & 5 (M:5; F:0) \\
  & Brain Station 23    & 5 (M:5; F:0) \\
\hline
\end{tabular}
\label{tab:demographics}
\end{table}

\subsection{Data Collection and Analysis}
We collected documents from institutional websites, government reports, and university portals. All materials were saved in .pdf format and organized into a structured archive. The data were analyzed using thematic analysis to identify curricular trends, policy references, and infrastructural indicators. This process generated 37 coded categories, which we sorted and grouped into clusters to identify the themes. They triangulated the findings with the interviews.

We also collected approximately six hours of audio recordings and 55 pages of field notes, which were transcribed and translated into English. We used a hybrid analytic approach combining inductive thematic coding with framework-informed analysis, following Braun and Clarke's six-phase approach \cite{braun2006using}. Transcripts from interviews were independently open-coded by two researchers using grounded theory techniques. This inductive process allowed themes to emerge directly from participant narratives rather than imposing predefined categories. The emergent codes were then clustered into higher-level themes, guided by dimensions from the \textit{aiEDU Readiness Framework} and the \textit{UNESCO AI Competency Framework}, including: curriculum integration, educator competency, ethics and responsibility, infrastructural capacity, and stakeholder engagement. To enhance validity and contextual depth, we triangulated these qualitative insights with document analysis of AI-related syllabus, course descriptions, faculty research profiles, and laboratory infrastructure descriptions collected in desktop research. Fourteen thematic codes naturally developed at the initial stage. These codes included themes such as "AI curriculum integration," "gender equity in AI," "industry-academia collaboration," "faculty competency," and "AI ethics." These codes were refined and grouped into broader themes, including "Curriculum and Faculty Competency," "Industry and Academic Partnerships," "Ethical AI and Societal Implications," and "Infrastructure and Resource Availability" (see supplementary materials for details). Coding disagreements between researchers were resolved through peer discussion. 


\subsection{Ethics and Positionality Statement}
The data collection was conducted by the researchers whose university does not mandate an IRB approval for this work. All participants were briefed on the study's goals and their rights, and informed consent was obtained prior to data collection. As all the authors of this paper are embedded in Bangladeshi higher education and the AI research ecosystem, we bring both proximity and positional bias. Our own experiences with AI instruction, student mentorship, and policy advocacy influence how we interpret these findings. At the same time, this positionality granted access to restricted institutional documents and trust-based recruitment pathways that would be otherwise difficult to achieve.


\section{Context}

Artificial Intelligence (AI) has emerged as a pivotal force in reshaping governance, education, and industrial development across the globe. Bangladesh’s national digital transformation agenda, articulated through the “Digital Bangladesh” Vision 2021 \cite{ged2015achieving}and subsequently extended under the “Smart Bangladesh 2041” framework \cite{pal2023smart}, positions artificial intelligence as a key driver of economic modernization, public service delivery, and global competitiveness. While structural challenges remain in infrastructure, human capital, and governance, Bangladesh is making marked efforts to align with global AI trends, including Responsible AI and ethical governance frameworks.

\subsection{AI Readiness Initiatives in Bangladesh}
Bangladesh's AI engagement formally took shape with the drafting of the National AI Strategy (2019–2024), signaling an early recognition of AI's transformative potential. Though implementation has been slow and key accountability mechanisms are still evolving, the strategy reflects an integrated vision that incorporates budget planning, stakeholder consultation, ethical considerations, and gender inclusion—particularly supporting women in AI and STEM fields \cite{ai_strategy_2024}. Parallel to the strategy, a draft National AI Policy was released in 2024 to guide the country's technological trajectory in alignment with international standards like UNESCO's Recommendation on the Ethics of Artificial Intelligence \cite{unescoethicsbd}.

The government's Telecommunication and ICT Ministry is the lead agency for AI governance, working closely with international bodies such as UNESCO \cite{ictd2025ai}. Through a series of consultation meetings and capacity-building efforts, Bangladesh has shown commitment to crafting a national approach that combines innovation with ethical safeguards. However, the draft policy is yet to be fully operationalized, and implementation frameworks, including monitoring indicators and evaluation loops, remain underdeveloped \cite{ai_strategy_2024}.

\subsection{Responsible AI and Ethical Governance Developments}
Bangladesh's foray into Responsible AI has been gradual but visible. While no formal national ethics framework for AI currently exists \cite{oxfordreadiness}, efforts are underway to embed principles of fairness, accountability, and transparency. Regulatory strides include the enactment of the Cyber Security Act (2023), development of a Data Protection Act (still pending enactment), and updates to older laws such as the Customs Act (2023) and the Consumer Protection Act (last amended in 2014) \cite{privacybd2023}. These legal tools provide partial scaffolding for ethical AI, especially in data privacy, cybersecurity, and consumer rights, though none explicitly address algorithmic bias, explainability, or AI safety.

Parallel to national policy development, Bangladesh has participated in promotional efforts to signal its interest in ethical AI practices. Initiatives like AI4Good summits, Digital Bangladesh Day exhibitions, and Startup Bangladesh events have featured AI as a core theme. However, these remain mostly symbolic and have not yet translated into formal regulatory tools or educational mandates aligned with global standards like UNESCO's \textit{Recommendation on the Ethics of Artificial Intelligence} \cite{unesco2023aiethics}. At present, Bangladesh does not have a national algorithmic impact assessment protocol or AI sandbox testing environment—features that are increasingly common in peer countries such as India and Indonesia \cite{oecd2023trackingai}. Additionally, Institutions like the Bangladesh e-Government Computer Incident Response Team (BGD e-Gov CIRT) and the National Cyber Security Agency are pivotal actors in mitigating cyber threats, laying the foundation for AI-related risk management \cite{bgdcirt}. Despite this progress, major gaps remain. Bangladesh still lacks clear accountability mechanisms for AI-related decisions, transparent algorithmic auditing procedures, and sector-specific ethical guidelines \cite{oxfordreadiness}. 

\subsection{AI Governance in Practice: Capacity and Limitations}


On the infrastructure side, Bangladesh has seen notable improvements under the ``Digital Bangladesh" initiative \cite{ged2015achieving}. It ranks 100th globally in the E-Government Development Index and 64th in broadband quality \cite{egdi2024}. Yet foundational challenges persist—ranging from limited GPU access and weak cloud computing capabilities to the lack of procurement transparency for emerging technologies like AI.

\subsection{Catching Up with Global Discourses}


Despite lacking a comprehensive legal framework for algorithmic governance, Bangladesh has shown interest in international models. The government has expressed intent to implement the UNESCO Ethics of AI Recommendation and has acknowledged sectoral prioritization for AI use in education, healthcare, and law enforcement \cite{unescoethicsbd}. Nonetheless, the public is not yet informed when AI systems make decisions about them—highlighting a key area for future transparency and citizen engagement \cite{unescoethicsbd}.

\section{Findings}
This section presents the key insights from our surveys, interviews, focus groups, and desktop research. 
It highlights the systemic gaps and limitations that currently inhibit effective AI readiness across Bangladeshi universities. 
We organize the findings under three themes: (a) material resources, (b) human resource and quality, and (c) curriculum and beyond. 
Each theme is further unpacked using four Science and Technology Studies (STS) concepts—\textit{blackboxing}, \textit{nested infrastructures}, \textit{strain}, and \textit{path dependency}—to situate the empirical accounts in a broader theoretical frame. 
We avoid naming subsections directly after the concepts; instead, we employ analytic titles grounded in the data while weaving in the theoretical engagement throughout. 
This layered approach allows us to show how local practices in Bangladesh reflect systemic processes of infrastructural inequality, historical lock-in, and sociotechnical tension.

\subsection{Material Resources}
A recurring theme across participants and institutional reviews was the scarcity of hardware, software, and infrastructural support for AI education. 
This scarcity was most vividly expressed through the lack of GPUs and high-performance computing infrastructure. 
As AI research becomes increasingly computationally intensive, relying on personal laptops or low-performance hardware severely limits students’ ability to engage with advanced deep learning or generative AI models. 
Without these resources, students often encounter AI as an abstract concept rather than a tangible practice. 
To unpack these dynamics, we analyze the data through four interrelated STS lenses.

\subsubsection{When AI Becomes Mathematics Alone}
Students frequently described their AI learning as reduced to mathematics without opportunities to experiment with models, datasets, or computational pipelines. 
This dynamic resonates with Latour’s notion of \textit{blackboxing}, in which the inner workings of a system become invisible and only inputs and outputs are observed. 
Without GPUs or hands-on labs, AI is blackboxed into formulas, leaving students to imagine rather than enact machine learning.

\begin{quote}
\textit{``During my undergrad at X university we covered the theoretical foundations of neural networks in depth, from activation functions to deriving the backpropagation algorithm by hand. But that was it. We never trained a model, never interacted with real datasets, and rarely used tools like TensorFlow or PyTorch. While we could explain the mathematics, we lacked the intuition that comes from seeing models learn and fail in practice.''} --- Student, X University
\end{quote}

The quote shows how students are asked to internalize AI conceptually while the infrastructure needed to ``open the box'' remains out of reach. 
This dynamic shapes not only their technical literacy but their very imagination of what AI is: abstract, formal, and disconnected from the messy realities of deployment. 
Industry accounts mirrored this problem. 

\begin{quote}
\textit{``I was assigned a task to develop an AI PoC for the company, but I couldn't meet the deadline because I had no experience working with GPUs in a real-world setting. I was used to working on personal laptops with minimal hardware, and when I tried running the models, the system couldn't handle the load. I ended up missing the project deadline, and the company had to bring in another developer with more experience. It was tough to get back on track.''} --- Junior AI Developer, Z Company
\end{quote}

Here, the blackboxing of computational infrastructure during education cascades into the workplace, where the unboxed reality of GPUs, cloud workflows, and scaling becomes painfully visible. 
This is not simply a gap in skills; it is the institutional production of ignorance through infrastructural exclusion.

\subsubsection{Global Markets in Local Labs}
Faculty and students consistently pointed out that their access to infrastructure was not simply a matter of local procurement but was nested in broader institutional, national, and global infrastructures. 
Star and Ruhleder’s work on \textit{nested infrastructures} emphasizes that infrastructures never stand alone; they are layered, interdependent, and shaped by wider political economies. 
Our findings illustrate this vividly.

\begin{quote}
\textit{``We don't have the funds to buy enough GPUs. AI is still comparatively new here, and the people in management who allocate funding are often not from computer science, so they don't fully understand why these resources are so critical. Space is another big problem — even though new buildings are being constructed, we haven't been allocated lab space. Politics plays a role too, older and more influential departments like social science or physics tend to get priority in funding and facilities, regardless of actual need. While top universities like IIT, NSU can provide decent internet and power backup, most of our PCs are too outdated to run modern deep learning models and would require a full upgrade.''} --- Faculty Member, X University
\end{quote}

This shows that GPU shortages are not just technical but the result of nested decision-making infrastructures: university administrators with disciplinary biases, national-level funding bodies like the UGC, and political hierarchies across departments. 
Moreover, global market forces shape which hardware is affordable or even available in Bangladesh. 

\begin{table*}[H]
\centering
\caption{Estimated GPU Availability and Student Population in AI-Focused Departments (2025)}
\begin{tabular}{|l|c|c|}
\hline
\textbf{University Department} & \textbf{Estimated Number of GPUs} & \textbf{Approx. Number of Students} \\
\hline
DU IIT & 2 & 148 \\
DU CSE & 5 & 240 \\
BUET CSE & 15--20 & 480 \\
BRACU CSE & 55 & 8,000 \\
NSU ECE & 39 & 6,000 \\
\hline
\end{tabular}
\label{tab:gpu_student_ratio}
\end{table*}

These figures reveal how infrastructures are unevenly distributed, not only within universities but between private and public institutions. 
BRACU and NSU, both private, have far more resources than DU or BUET, which are public. 
This reflects the nested infrastructure of educational inequality in Bangladesh: private wealth allows hardware investment, while public institutions remain dependent on bureaucratic and donor logics.

\subsubsection{The Weight of Scarcity}
The accounts of students and faculty also illustrate the concept of \textit{strain}, developed by Strauss to describe the tension when social worlds are stretched beyond their capacities. 
In Bangladesh, AI education promises global relevance but delivers without sufficient resources, producing widespread frustration and stress.

\begin{quote}
\textit{``I couldn't complete my project on time because I didn't have access to a decent system or enough storage to run models. It's frustrating when you have the theoretical knowledge but can't test or deploy it because the tools simply aren't available. It's a huge barrier, and it feels like we're being set up for failure when the infrastructure isn't there to support us.''} --- Student, X University
\end{quote}

This sense of being ``set up for failure'' captures the experiential weight of strain. 
Faculty echoed similar frustrations, balancing heavy teaching with resource deficits. 
Industry leaders saw this in graduates as well: the junior developer’s story above illustrates how strain is not only academic but extends into early professional life, where underprepared graduates struggle to meet organizational expectations. 
Strain here becomes systemic: a stress fracture running across the pipeline from education to industry.

\subsubsection{Locked-in Priorities}
The distribution of material resources is marked by \textit{path dependency}. 
Because AI was historically peripheral to Bangladesh’s national and institutional agendas, it continues to be deprioritized in budgets and policies. 
Once these trajectories are set, they become self-reinforcing, even in the face of changing global conditions.

GPU allocation follows institutional histories: older departments dominate funding, while computer science struggles to compete. 
This reflects what Pierson and Arthur describe as ``lock-in'' effects \cite{arthur1994increasing, pierson2000increasing}: once institutional rules and norms are established, they bias future decisions, even when alternatives are clearly superior. 
The very process of resource allocation is path-dependent, ensuring that deficits today will reproduce deficits tomorrow.

\subsection{Human Resource and Quality}
If material infrastructures constitute the visible bottlenecks of AI education, then human resources—faculty, researchers, students, and their networks—represent the less visible but equally consequential dimensions of readiness. Our findings reveal multiple, interlocking challenges around the preparation, training, and equitable participation of human actors. These include a lack of formal and informal training opportunities for faculty and students, minimal alumni and industry engagement, entrenched gender disparities, and professional cultures that reproduce exclusionary norms. To analyze these findings, we mobilize four STS concepts—blackboxing, nested infrastructures, strain, and path dependency—framed through empirically grounded subsubsections.

\subsubsection{Teaching Without Training (Strain)}
Across institutions, faculty emphasized the absence of structured opportunities for upskilling in AI. With no formal mechanism for continuous professional development, they relied on self-study, occasional workshops, or free online courses. This created a paradox: faculty were expected to deliver cutting-edge instruction while being systematically denied the resources to keep pace with the field. This exemplifies Strauss’s notion of \textit{strain}, in which institutional demands exceed available capacities, producing stress fractures in practice.

\begin{quote}
\textit{``I often try to keep my students updated with the latest trends in AI, but without formal training or opportunities to collaborate with global experts, there's a significant gap in what we can provide. The lack of institutional support for international exposure and ongoing professional development creates a barrier to advancing both faculty and student knowledge.''} --- Faculty, Y University
\end{quote}

The consequences of these gaps were visible in research output. BUET, with the highest cumulative citation count (21,768), still averaged only 1,281 citations per faculty. BRACU, by contrast, had 10,476 total citations with an average of 582 per faculty. These figures lag significantly behind global peers, underscoring how institutional neglect of professional development shapes research visibility.

\begin{figure}[!t]
\centering
\includegraphics[width=0.7\columnwidth]{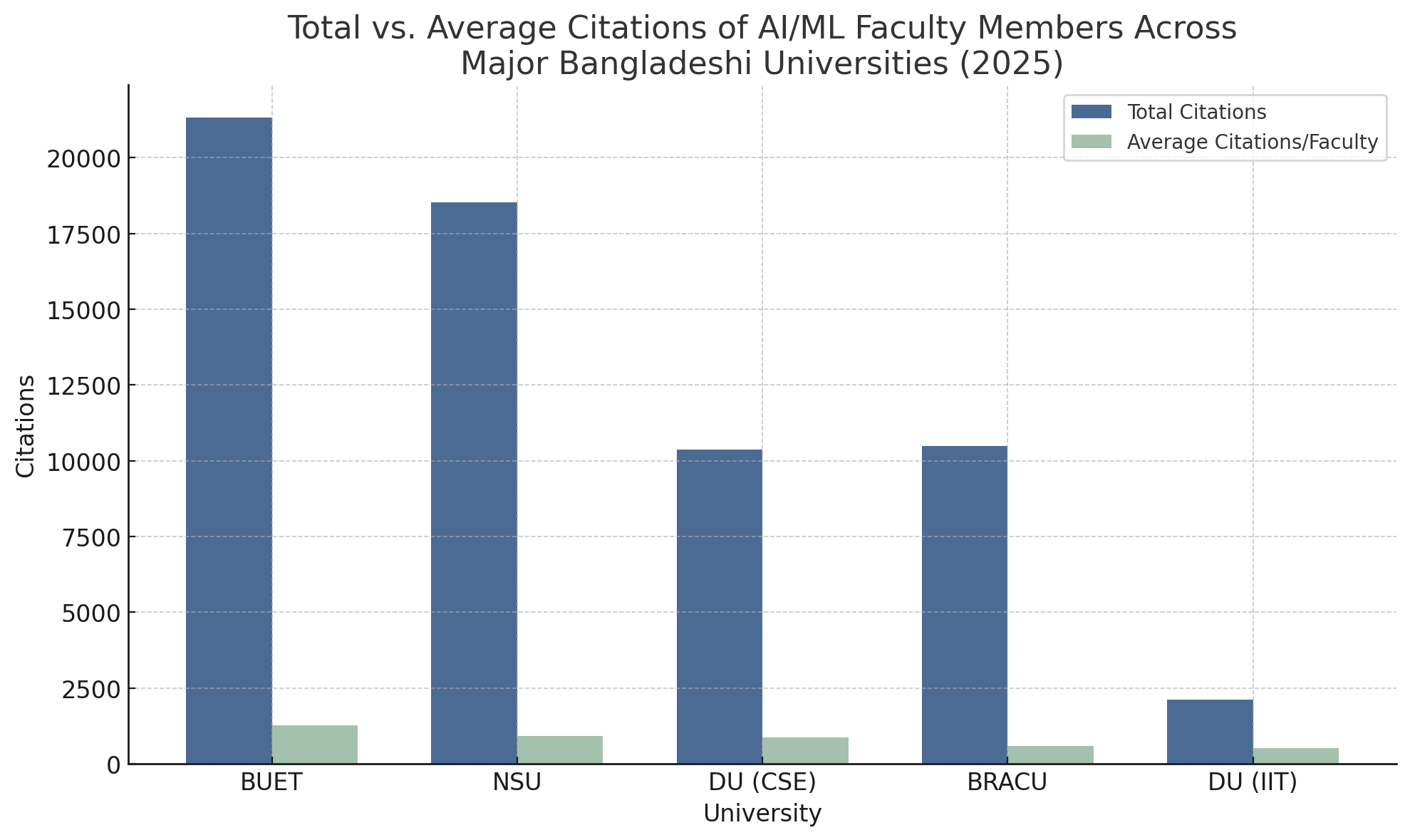}
\caption{Cumulative Citation Analysis of Total Citation vs Average Citation per faculty in AI/ML fields in Bangladeshi Universities}
\label{fig:citations_chart}
\end{figure}

STS scholarship on repair and breakdown \cite{jackson201411} helps explain this paradox. Faculty often engaged in “workarounds,” cobbling together knowledge from free resources while under pressure to present themselves as up-to-date educators. This repair work is invisible in official accounts of readiness but critical for sustaining AI education under scarcity. The strain emerges because repair cannot keep pace with the rapid evolution of global AI, leaving both faculty and students behind.

The situation also reflects how infrastructures are political. Decisions not to invest in faculty training reproduce inequities between institutions: elite universities with international collaborations occasionally access training, while others remain locked out. This unevenness creates a stratified human infrastructure for AI education, where readiness is concentrated in pockets but absent across the majority of institutions.

\subsubsection{Mentorship as Privilege (Blackboxing)}
Students frequently reported that access to mentorship, research groups, and informal learning opportunities depended on insider connections, CGPA, or sheer luck. These accounts illustrate \textit{blackboxing} at the level of mentorship: the pathways into AI research are treated as if they “just exist,” while the social labor and exclusions that sustain them remain invisible. For those on the outside, the system appears closed, opaque, and unattainable.

\begin{quote}
\textit{``AI research groups exist, but access often depends on your connections and mostly on the basis of initial year's CGPA. If you don't have a high CGPA or you're not part of the right network, you can easily miss out on valuable learning opportunities, mentorship, and projects. It shouldn't be this way. AI should be accessible to anyone passionate about it, regardless of who they know.''} --- Student, Z University
\end{quote}

In STS terms, mentorship operates like an infrastructure: it is most visible when it breaks down or when one is excluded from it. For students outside key networks, the absence of access reveals the underlying gatekeeping practices. Bowker and Star’s work on classification is useful here \cite{bowker2000sorting}: students are sorted and classified early (by GPA or social ties), and these classifications shape long-term trajectories. Once classified as “outside” the mentorship circuit, it becomes difficult to re-enter.

Alumni and industry professionals confirmed this from the other side. One engineer explained:

\begin{quote}
\textit{``Honestly, it's not happening because no one is taking initiative. Faculty aren't reaching out, and engineers who came from these institutions aren't going back to help. Everyone's too comfortable in their own bubble.''} --- AI Engineer, Y Company
\end{quote}

Another noted:

\begin{quote}
\textit{``Universities have talented alumni working in AI globally. If even a few took the initiative to host regular seminars or guest lectures, students would be far more in tune with current industry trends. But right now, that kind of engagement is practically nonexistent.''} --- AI Engineer, Z Company
\end{quote}

Here, the blackboxing of mentorship extends to alumni engagement. Knowledge flows are assumed but rarely materialize. The relational infrastructures that could connect alumni to students are invisible, unaccounted for, and thus inactive. Students see only the absence of opportunities, not the hidden labor and institutional inertia that prevent engagement.

For HCI, these dynamics matter because mentorship is a crucial site where norms, values, and practices of a field are reproduced. When mentorship is blackboxed and privatized, it reinforces inequity, limiting who gets to shape the future of AI in Bangladesh.

\subsubsection{Networks of Exclusion (Nested Infrastructures)}
Gender disparities emerged as the most striking form of exclusion. Faculty rosters across surveyed institutions were overwhelmingly male: NSU’s ECE had only 3 female faculty out of 20, BRACU had 2 out of 18, BUET had 4 out of 17, and DU’s CSE and IIT departments had none. This absence of women in academic leadership deprives students of role models and reinforces perceptions of AI as male-dominated.

\begin{figure}[!t]
\centering
\includegraphics[width=0.7\columnwidth]{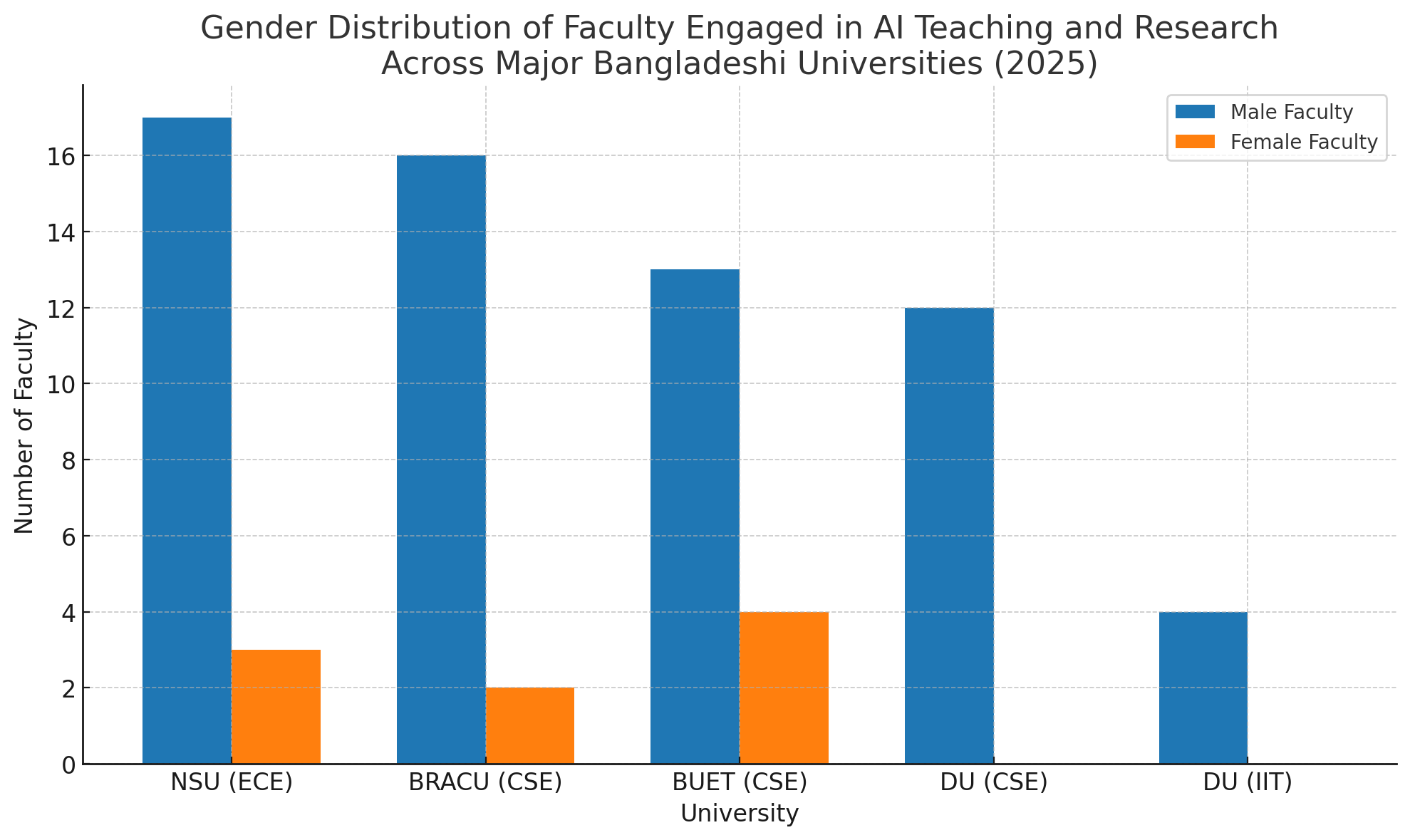}
\caption{Gender Distribution of Faculty Engaged in AI Teaching and Research Across Major Bangladeshi Universities}
\label{fig:gender_chart}
\end{figure}

Female students described how exclusion operated not only structurally but interpersonally:

\begin{quote}
\textit{``As a female student at Z University, I often feel like I have to prove myself twice as much during the capstone project selection process. Even when I take the lead and contribute significantly, the credit is often unfairly attributed to my male teammates. This not only undermines our efforts but also perpetuates the idea that women aren't as capable of taking charge in technical fields, despite the reality of our work and leadership.''} --- Female Student, Z University
\end{quote}

\begin{quote}
\textit{``Internship information usually spreads through boys' groups on WhatsApp. We're not always in those loops.''} --- Female Student, P University
\end{quote}

These accounts illustrate how exclusion is produced through \textit{nested infrastructures}. Gender disparity is not only about hiring policies but also about overlapping systems: faculty recruitment norms, peer communication practices, and industry hiring cultures. Together, these infrastructures reproduce exclusion at multiple scales.

\begin{figure}[!t]
\centering
\includegraphics[width=0.7\columnwidth]{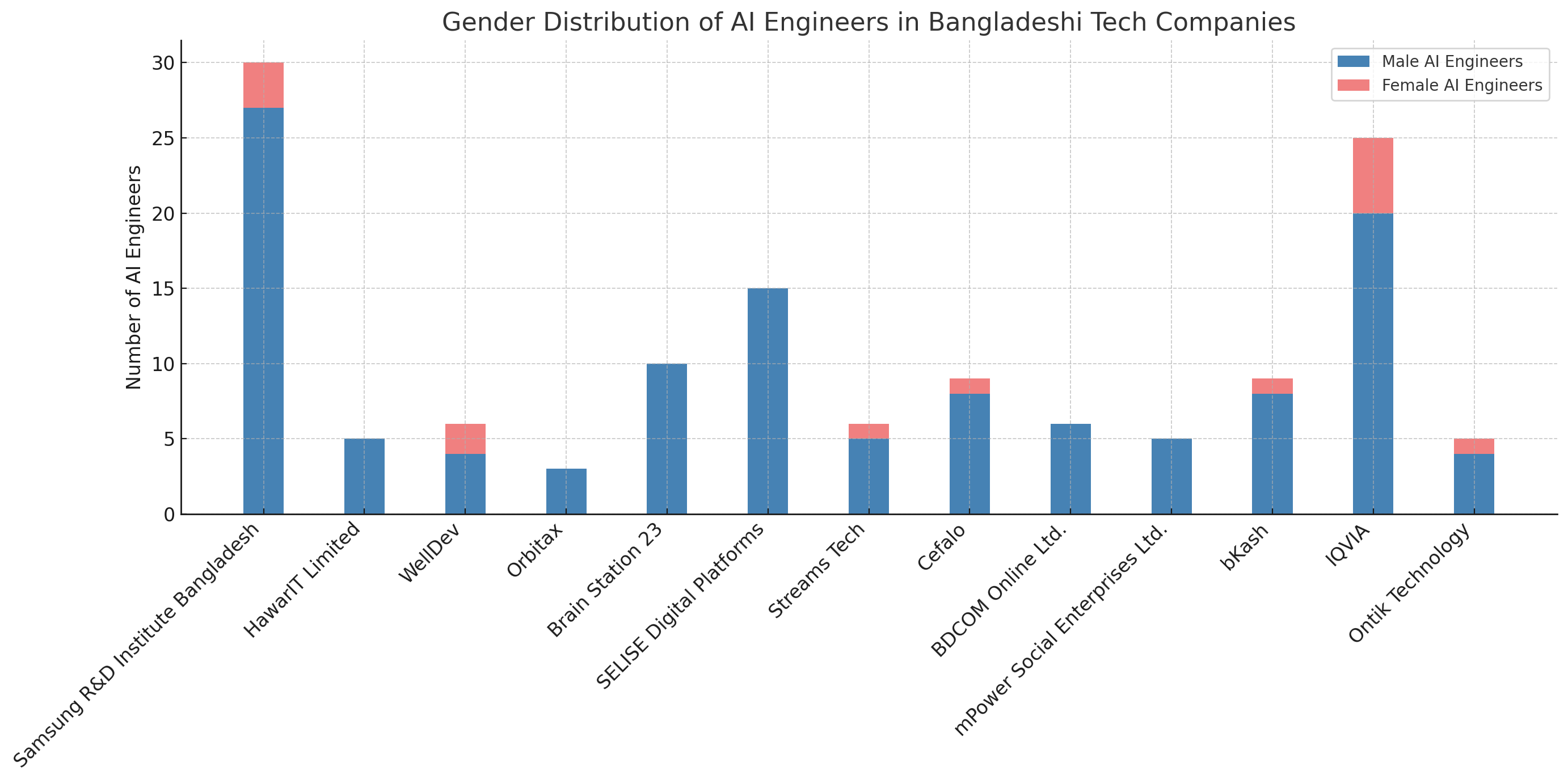}
\caption{Gender Distribution of AI Engineers in Leading Bangladeshi Tech Companies}
\label{fig:industry_gender_chart}
\end{figure}

Industry mirrors these academic disparities. Companies such as Samsung and IQVIA employ very few women in AI roles, as illustrated in Figure~\ref{fig:industry_gender_chart}.

Feminist STS scholarship by Suchman and  Wajcman helps us see this not as incidental but as structural \cite{suchman1987plans, wajcman2004technofeminism}. Technologies and their institutions are gendered: norms of who belongs and who leads are built into hiring, training, and even informal WhatsApp groups. The nested infrastructures of academia and industry reinforce one another, ensuring that exclusion persists despite individual talent or motivation.

\subsubsection{Cycles of Inequality (Path Dependency)}
Gender exclusion and weak faculty training also demonstrate \textit{path dependency}. Historical underrepresentation of women in STEM creates a cycle: because women are absent in faculty roles, students lack role models, which discourages future cohorts, perpetuating male dominance. Similarly, institutional undervaluing of research has locked universities into teaching-heavy models, making it difficult to build the research cultures necessary for AI readiness.

As Pierson and Arthur note, path dependency occurs when early institutional decisions create self-reinforcing mechanisms \cite{arthur1994increasing, pierson2000increasing}. In Bangladeshi AI education, the failure to invest in equitable hiring and robust faculty development has created long-term inertia. Even when faculty and administrators recognize the problem, the structural momentum of past practices constrains what can be done. This explains why gender disparities persist despite recognition of their harms and why research outputs remain low even as global AI advances.

Students also see the consequences of path dependency in mentorship:

\begin{quote}
\textit{``AI research groups exist... If you don't have a high CGPA or you're not part of the right network, you can easily miss out on valuable learning opportunities, mentorship, and projects.''} --- Student, Z University
\end{quote}

Once excluded early, it is difficult to break back in. Path dependency here is not only institutional but biographical: trajectories are shaped in the first year, and opportunities compound from there.

\subsection{Curriculum and Beyond}
Curricula reflect how knowledge itself is shaped, stabilized, and constrained. 
Our findings reveal outdated syllabi, limited interdisciplinarity, and the absence of Responsible AI and ethics training. 
These are not merely pedagogical shortcomings but manifestations of sociotechnical processes: knowledge is blackboxed into abstractions, reforms are stalled within nested bureaucracies, tensions between theory and practice create strain, and institutional inertia produces path-dependent stagnation. 
We unpack these dynamics through four interrelated lenses.

\subsubsection{Syllabi as Sealed Boxes}
Students described AI education as reduced to theory, with little practical exposure. 
This resonates with Latour’s notion of \textit{blackboxing}: the complex work of model building and failure remains invisible, leaving only abstract derivations as inputs and grades as outputs.

\begin{quote}
\textit{``During my undergrad at X university we covered the theoretical foundations of neural networks in depth, from activation functions to deriving the backpropagation algorithm by hand. But that was it. We never trained a model, never interacted with real datasets, and rarely used tools like TensorFlow or PyTorch. While we could explain the mathematics, we lacked the intuition that comes from seeing models learn and fail in practice.''} --- Student, X University
\end{quote}

This blackboxing of curricula shapes how students imagine AI: as equations, not as a socio-technical practice. 
Industry leaders echoed this disconnect. One engineer explained:

\begin{quote}
\textit{``I interviewed (for job) a student from one of the country's top public universities and asked, `Why use AI here? It can be solved in another way.' He couldn't answer. He knew the tools but not the reasoning behind them. That's where our curriculum is failing. Students aren't learning to solve real-world problems. Sure, you can connect four pieces of wood and call it a structure but unless they're aligned with purpose, it won't stand. Our AI education still revolves around outdated projects like chess bots, instead of tackling meaningful, real-world challenges. We're falling behind.''} — Lead Engineer, X Company
\end{quote}

Here, the blackboxing is not incidental but structured: curricula conceal the messy reasoning behind applied AI, privileging technical recall over critical judgment.

\subsubsection{The Bureaucracy of Reform}
Faculty noted that updating syllabi was hindered by multiple layers of approvals, especially from the University Grants Commission (UGC). 
This illustrates \textit{nested infrastructures}: local curricular practices are embedded in institutional and regulatory layers that constrain change.

\begin{quote}
\textit{``The process of changing the curriculum is incredibly time-consuming. You need multiple approvals from the UGC and the academic board, which means any proposed updates, especially those related to new topics like AI ethics, can take months—sometimes even years. By the time the curriculum is updated, it's often already outdated.''} --- Faculty, Z University
\end{quote}

These bureaucratic delays mean global frameworks like UNESCO’s AI ethics recommendations rarely translate into practice. 
Table~\ref{tab:ethics_courses} illustrates the absence of AI ethics across major universities.

\begin{table*}[H]
\centering
\caption{AI Ethics Course Availability in Bangladeshi Universities (2025)}
\begin{tabular}{|l|l|l|}
\hline
\textbf{University} & \textbf{AI Ethics Course Offered} & \textbf{Remarks} \\
\hline
NSU (ECE) & General (PHI 104) & Covers broad ethics; no AI-specific content \\
BRACU (CSE) & General (HUM 347) & Discusses tech and society; lacks AI focus \\
BUET (CSE) & None & No ethics or Responsible AI courses \\
DU (CSE) & None & No coverage of AI ethics, fairness, or safety \\
DU (IIT) & None & Entirely technical syllabus; ethics not mentioned \\
\hline
\end{tabular}
\label{tab:ethics_courses}
\end{table*}

The absence of ethics courses is thus not accidental but produced by nested bureaucracies: each layer slows responsiveness, reproducing outdated content despite awareness of ethical risks.

\subsubsection{Tensions Between Theory and Practice}
The tension between classroom theory and industry practice exemplifies Strauss’s notion of \textit{strain}. 
Students are trained in mathematical derivations but judged against industry expectations of hands-on problem-solving. 
Faculty themselves recognized this gap:

\begin{quote}
\textit{``Unfortunately, we are not fully updated with the current trends in industry. This is partly due to a lack of interest from both sides and partly due to ego issues — industry often thinks academia does not understand practical constraints, while academia believes it is engaged in 'real' research. In other countries, collaborations are common, but here in Bangladesh they are rare. Even when both sides want to collaborate, factors like city traffic, long commute times, and workload make it difficult. We have revised our AI syllabus with industry feedback in the past two years, but guest lectures or hands-on training sessions from industry experts have become rare.''} --- Faculty Member, X University
\end{quote}

Students confirmed the absence of ethics and fairness:

\begin{quote}
\textit{``I've never been taught about data privacy or fairness in any AI course. We just focus on math and code.''} --- Student, X University
\end{quote}

The consequences extend into industry. A senior manager recounted:

\begin{quote}
\textit{``Yes, actually — we had a junior developer once, fresh out of university, who built an essay scoring model for a government project. Technically, it was impressive. But he ignored fairness, used real student data without consent, and didn’t think twice about explainability. The model ended up disadvantaging rural students and leaking personal info. It cost the company a major contract — and it cost him our trust. That's when we learned: in AI, ethical shortcuts always come with a price.''} — Senior Manager, X Company
\end{quote}

These accounts underscore how strain is not only academic but societal: curricula promise readiness but deliver graduates unequipped for ethical or applied challenges.

\subsubsection{Chess Bots Forever?}
The persistence of outdated content, such as classical projects like chess bots, reveals \textit{path dependency}. 
Once curricular trajectories are established, they reproduce themselves through institutional inertia, even when globally obsolete.

\begin{table*}[t!]
\centering
\resizebox{\textwidth}{!}{%
\begin{tabular}{|p{3.6cm}|p{2.8cm}|p{2.8cm}|p{2.8cm}|p{2.8cm}|p{2.8cm}|p{2.8cm}|}
\hline
\textbf{Dimension} & \textbf{Top Global Programs} & \textbf{DU IIT } & \textbf{DU CSE} & \textbf{BracU} & \textbf{NSU} & \textbf{BUET} \\
\hline
\textbf{Human-Centred Mindset} & Strong focus on integrating societal impacts, ethical AI, and user-centered design. & No focus on human-centred mindset. & Rarely focuses on human-centred design, mostly on technical aspects. & Human-centred focus is mentioned informally, but not deeply embedded. & Minimal coverage of human-centred aspects, lacks in-depth discussions on the impact of AI. & Not mentionable focus on human-centred mindset, little integration of societal impacts. \\
\hline
\textbf{Ethics of AI} & Dedicated courses on AI ethics, societal impacts, and responsible AI. & No dedicated course on AI ethics. & No dedicated course on AI ethics. & No dedicated course on AI ethics. & No dedicated course on AI ethics. & No dedicated course on AI ethics. \\
\hline
\textbf{AI Techniques and Applications} & Coverage of cutting-edge AI techniques, hands-on exposure to modern tools (e.g., LLMs, generative AI). & Coverage of AI concepts is theoretical, limited hands-on experience with modern tools. & Theoretical coverage of AI techniques with limited practical exposure. & Basic understanding of AI techniques with minimal real-world application. & AI techniques are covered at a basic level, but advanced topics like LLMs or generative AI are rarely discussed. & Covers AI techniques theoretically but lacks modern practical implementation. \\
\hline
\textbf{AI System Design} & Strong industry collaborations, capstone projects with real-world impact, mentorship opportunities. & Capstone projects exist, but lack significant real-world impact or mentorship. & Limited AI system design, with minimal project-based work. & Capstone projects are more academic, with no major focus on AI system design. & Limited practical work, mostly theoretical concepts with a lack of industry-relevant design work. & Very limited AI system design work; mostly academic theoretical exercises without real-world application. \\
\hline
\end{tabular}
}
\caption{Global AI Curriculum vs. Bangladeshi AI Curriculum based on UNESCO AI Competency Framework for Students (2025)}
\end{table*}

As our benchmarking matrix shows, Bangladeshi universities lag behind global exemplars in integrating ethics, hands-on system design, and modern AI tools. 
Faculty acknowledged efforts to revise syllabi, but bureaucratic cycles and departmental rivalries ensure that legacy topics endure. 
This reflects classic lock-in: institutional pathways reinforce themselves, marginalizing reform.

Curricula in Bangladesh demonstrate how knowledge infrastructures can become self-perpetuating. 
Syllabi blackbox AI into abstractions, bureaucratic nesting slows reform, tensions between theory and practice generate strain, and path dependency locks outdated projects into place. 
Curriculum here is not just a pedagogical artifact but an infrastructure that reproduces inequality and stagnation, leaving students underprepared for both ethical and applied dimensions of AI.

\section{Discussion}
This study of AI readiness in Bangladeshi universities contributes to HCI and ICTD scholarship by demonstrating how readiness is not only a matter of material resources or curriculum design but a deeply sociotechnical condition. By analyzing empirical findings through the lenses of blackboxing, nested infrastructures, strain, and path dependency, we reveal how infrastructural scarcity, opaque mentorship, institutional stress, and historical lock-in shape the everyday realities of AI education. We also extend these analyses by situating AI readiness within postcolonial critiques of global computer science education, highlighting how epistemic dependency and market logics reinforce inequality. Our approach contributes to longstanding HCI debates on infrastructure \cite{star1994steps, bowker2000sorting}, breakdown and repair \cite{jackson201411}, feminist HCI \cite{suchman1987plans, wajcman2004technofeminism}, and postcolonial computing \cite{127irani2010postcolonial}. In what follows, we organize our discussion into five themes. Each subsection elaborates theoretical contributions to HCI and ICTD, while offering practical implications for design, policy, and pedagogy.

\subsection{Implications for Design and Policy}

Our findings highlight several problems that demand immediate, design-oriented responses. First, the blackboxing of mentorship into opaque networks suggests the need for lightweight, transparent platforms that match students with faculty and alumni based on interest areas rather than GPA or personal connections. Second, the absence of Responsible AI instruction could be addressed through modular online toolkits or plug-in course units, allowing faculty to introduce case studies on fairness, privacy, and accountability without waiting for full curriculum reform. Third, the scarcity of GPU access can be mitigated in the short term by leveraging cloud-based credits and cooperative lab models, where institutions pool resources to create shared training environments. These solutions do not resolve systemic inequities, but they provide immediate, actionable interventions that could enhance visibility, access, and ethical literacy in AI education. They align with HCI’s tradition of prototyping: building partial but practical fixes that de-blackbox practice and distribute opportunities more equitably.

Other problems identified in our findings, however, demand more structural and policy-level interventions. The persistence of outdated syllabi, for example, reflects path dependency rooted in UGC approval cycles and departmental politics; breaking this lock-in requires reform of governance structures and participatory curriculum design involving students, industry, and faculty. Gender disparities in both academia and industry cannot be solved by isolated mentoring programs; they call for infrastructural change in hiring practices, funding for women’s research groups, and institutional policies that address bias in evaluation and leadership. Finally, the lack of interdisciplinary collaboration across domains such as health, agriculture, and law underscores the need for cross-sectoral infrastructure: shared data platforms, joint labs, and national research networks that incentivize collaboration. These problems are not amenable to immediate design fixes; they require infrastructural investments and policy development that recognize AI readiness as a long-term, systemic project. For HCI and ICTD, the implication is clear: interventions must operate on two timescales—short-term design innovations that alleviate immediate inequities, and long-term structural reforms that reconfigure the sociotechnical infrastructures of education and innovation.

\subsection{Rethinking AI Readiness as Sociotechnical Infrastructure}
Our findings show that AI readiness cannot be reduced to a checklist of hardware, curricula, or policy aspirations. Instead, readiness is a layered sociotechnical infrastructure, with material, human, and curricular dimensions that are deeply interdependent. Star and Ruhleder’s notion of \textit{nested infrastructures} helps us see why GPU scarcity is not simply a procurement issue but entangled with bureaucratic politics, donor logics, and global supply chains \cite{star1994steps}. Similarly, faculty development cannot be understood in isolation: it is embedded within institutional funding priorities, international collaborations, and professional cultures that devalue continuous training. Infrastructural inversion, as Bowker and Star elaborate, allows us to see how invisible background systems—such as university politics, approval processes, and budget allocations—shape everyday student experiences \cite{bowker2000sorting}. What looks like “resource shortage” on the surface is in fact the emergent outcome of layered infrastructural dependencies.

This reframing has implications for HCI research that often treats AI adoption in Global South contexts as a matter of “capacity gaps.” By conceptualizing readiness as infrastructural, we move away from deficit narratives and toward an understanding of readiness as an ecology of interdependencies. Such a shift resonates with work on ICTD infrastructures \cite{rangaswamy2011cutting, best2007gender}, which shows how infrastructural fragility and improvisation are defining features of technology use. The concept of infrastructure also foregrounds repair \cite{jackson201411}: the hidden work of faculty who self-update with MOOCs, or students who crowdsource cloud credits, is infrastructural repair in the face of scarcity. Recognizing these practices challenges technocratic visions of readiness that assume smooth pipelines of investment and policy.

For design, this reframing suggests that interventions must target ecologies rather than isolated fixes. Providing GPUs without reforming budget processes may not alleviate shortages. Revising syllabi without addressing bureaucratic inertia may not yield meaningful reform. Readiness must be treated as a sociotechnical assemblage, requiring infrastructural investments, new governance models, and collaborative practices across academia, industry, and government. For HCI, this contribution lies in extending infrastructural analysis from end-user systems to the educational and institutional infrastructures that determine who gets to participate in AI at all. Therefore, we argue that AI readiness must be redefined in HCI as an infrastructural condition—requiring analysis and design at the level of ecologies, not isolated technological fixes.

\subsection{Blackboxing and its Consequences for Pedagogy and Practice}
Blackboxing provides a useful analytic for understanding how AI education in Bangladesh remains opaque at multiple levels. In Latour’s original sense \cite{latour1987science}, blackboxes emerge when systems work smoothly and their inner workings fade from view. In Bangladeshi AI curricula, blackboxing emerges differently: through absence. Students memorize derivations without ever “opening the box” of model training, dataset curation, or iterative debugging. This is a form of \textit{closed worlds}, contexts in which the inner workings of systems are hidden, naturalized, or rendered inaccessible. When students are denied computational resources, blackboxing is pedagogically enforced, producing a vision of AI as pure mathematics divorced from messy practice.

Mentorship, too, is blackboxed. Students described access to research groups as dependent on opaque criteria: GPA rankings, personal networks, or sheer chance. Alumni described a lack of initiative on both sides—faculty not reaching out, industry not investing time. These conditions render the infrastructure of mentorship invisible: students do not see the labor, politics, and exclusions that produce access. For HCI, this extends Latour’s concept from technical systems to educational systems: pedagogy itself can function as a blackbox, obscuring the sociotechnical work behind participation.

The consequences are significant. Students graduate with technical vocabulary but lack the reasoning skills to apply them. As our industry respondents noted, graduates “know the tools but not the purpose.” This opacity shapes epistemic cultures \cite{cetina1999epistemic}: what counts as knowledge is narrowed to derivations, while reasoning, application, and ethics are excluded. The result is not only skill gaps but epistemic injustice: students from less-connected networks cannot access mentorship, and women are excluded from informal WhatsApp loops. Blackboxing thus reproduces inequity.

For HCI, the implication is clear: we must design to de-blackbox. Pedagogical platforms can make failures visible, allowing students to see where models break. Mentorship infrastructures can be formalized to reduce reliance on opaque networks. Open datasets, collaborative coding environments, and transparency in evaluation criteria can all serve to de-blackbox AI education. The contribution here is conceptual and practical: extending blackboxing from technical systems to educational infrastructures opens new design spaces for equity and transparency. Therefore, we argue that HCI must treat educational blackboxes with the same critical scrutiny as algorithmic ones, designing interventions that make hidden practices visible and equitable.

\subsection{Strain, Inequality, and the Human Costs of Readiness}
Our study illustrates Strauss’s concept of strain: the tensions that arise when institutional expectations exceed available resources. Faculty are expected to deliver globally competitive curricula without access to training or computing infrastructure. Students must navigate opaque mentorship systems and short academic calendars, producing a sense of being “set up for failure.” Women experience compounded strain, needing to overperform to gain recognition while also being excluded from informal networks where opportunities circulate. Strain reveals not just material shortages but human costs: exhaustion, disillusionment, and inequity.

STS scholarship on breakdown and repair shows how strain is often absorbed by hidden repair work \cite{jackson201411}. Faculty repair institutional shortcomings by self-teaching; students repair mentorship gaps by hustling for connections. These repairs sustain the system but also mask the depth of dysfunction, preventing systemic reform. For women, repair often takes the form of overcompensation: leading projects twice as effectively to gain half the recognition. These practices echo feminist HCI accounts of invisible labor \cite{d2023data}, showing how readiness is sustained by unequal human effort.

Strain also reflects what Suchman calls situated action \cite{suchman1987plans}: readiness plays out in specific contexts of classrooms, labs, and job interviews, not in abstract policy documents. For example, the female student excluded from WhatsApp internship groups experiences strain not as a policy gap but as an everyday injustice. This illustrates why HCI must move beyond macro-policy metrics of readiness to the lived experiences of students and faculty.

Design implications include creating support systems that alleviate strain rather than amplify it. Professional development platforms could provide structured training for faculty. Mentorship matching systems could reduce reliance on GPA or networks. Gender-sensitive policies could ensure equitable distribution of opportunities. For HCI, the contribution is theoretical and practical: reframing readiness as lived strain highlights the importance of designing not just for efficiency but for equity and care in educational infrastructures. Therefore, we argue that readiness metrics must include the human costs of strain, positioning well-being and equity as central outcomes of sociotechnical design.

\subsection{Path Dependency and Futures of Responsible AI}
The persistence of outdated syllabi, the dominance of theory-heavy teaching, and the exclusion of ethics all reflect the path-dependent nature of AI education. Once trajectories are set—chess bots as canonical projects, mathematics as the sole arbiter of expertise—they reproduce themselves through institutional inertia. Pierson explains this through increasing returns: the longer a path is followed, the more costly it becomes to deviate \cite{pierson2000increasing}. In Bangladeshi AI curricula, once syllabi are approved by the UGC, they remain locked in for years, even as the field evolves rapidly. This institutional lock-in is compounded by departmental rivalries and ego-driven conflicts, ensuring that reforms are piecemeal and slow.

Path dependency explains why Responsible AI is absent despite global emphasis. Once ethics is excluded from the curriculum, there are few incentives to add it later: faculty are untrained, syllabi are full, and industry does not demand it. This creates what Arthur calls technological lock-in \cite{arthur1994increasing}: inferior practices persist because switching costs are high. The result is a generation of engineers trained without exposure to fairness, privacy, or accountability. As one industry manager noted, the absence of ethical training led to real-world harm when a model disadvantaged rural students.

For HCI, this raises a critical design question: how do we intervene in path-dependent systems to enable new futures? One approach is infrastructural seeding: creating pilot ethics courses or cross-sector labs that demonstrate value and create pressure for adoption. Another is participatory curriculum co-design, involving students, industry, and faculty in building new trajectories. International partnerships can also disrupt inertia by embedding ethics into joint programs. The contribution here is theoretical—extending path dependency to educational systems—and practical: offering pathways to break lock-in and cultivate Responsible AI futures. Therefore, we argue that breaking curricular path dependencies is an urgent design challenge for HCI, demanding participatory and cross-sector interventions that seed new trajectories.

\subsection{Postcoloniality in Computer Science Education}
Finally, our findings point to the need for a postcolonial lens on AI education in Bangladesh and the Global South. Curricula, benchmarks, and even hiring expectations are overwhelmingly imported from Euro-American contexts, often without adaptation to local realities. This reproduces what Ahmed \cite{ahmed2024living} and Irani et al. describe as epistemic dependency \cite{127irani2010postcolonial}: local expertise is framed as derivative rather than generative. The market dimension compounds this problem. Multinational corporations demand skillsets aligned with global platforms (TensorFlow, Hugging Face) while undervaluing local applications in agriculture, health, or governance. Students thus learn to optimize for global labor markets rather than local social challenges.

Infrastructural deficits—scarce GPUs, limited faculty training—become further evidence of “lagging behind,” reinforcing deficit narratives of the Global South. Postcolonial STS \cite{harding1998science, anderson2002introduction} reminds us that knowledge production is unevenly distributed, with Global South actors positioned as users rather than producers of innovation. Feminist postcolonial computing Irani et al. urge us to foreground local epistemologies and resist reproducing colonial hierarchies in design \cite{127irani2010postcolonial}. For AI readiness, this means asking whose futures are being prepared: are students being trained to serve global markets, or to address local inequities?

For HCI, adopting a postcolonial lens means resisting epistemic dependency. This includes designing curricula that integrate local data, applications, and ethical concerns. It requires recognizing the political economy of AI labor, where Bangladeshi graduates may serve as annotation workers for global systems rather than innovators of local solutions. It also requires reframing readiness away from deficit comparisons to the Global North and toward contextually grounded innovation. The contribution here is critical: highlighting how AI readiness is not just infrastructural but geopolitical, shaped by postcolonial dynamics of knowledge and labor. Therefore, we argue that HCI must adopt a postcolonial stance in AI education, resisting epistemic dependency and re-centering local knowledges, markets, and priorities in the Global South.

Across these five themes, our discussion demonstrates how AI readiness in Bangladesh is best understood as a sociotechnical, infrastructural, and postcolonial condition. Blackboxing, nested infrastructures, strain, and path dependency reveal how inequities are reproduced not only through resource scarcity but through epistemic and institutional practices. Our postcolonial lens highlights how global market logics and Euro-American benchmarks reinforce dependency, framing local education as derivative. For HCI, the contributions are fourfold: (1) reframing readiness as infrastructural, (2) extending blackboxing to educational systems, (3) theorizing readiness as lived strain, and (4) conceptualizing curricula as path-dependent systems that demand new design interventions. Together, these contributions advance HCI and ICTD debates on infrastructure, equity, and decoloniality, while offering actionable implications for the design of curricula, mentorship infrastructures, and professional development platforms. Ultimately, AI readiness must be understood not as catching up to a global benchmark but as cultivating infrastructures of equity, care, and responsibility in local contexts.

\section{Limitations}
While our study provides one of the most detailed examinations of AI readiness in Bangladeshi higher education, several limitations must be acknowledged. First, our data is based on a purposive sample of universities, industry stakeholders, and policymakers, rather than a nationally representative survey. This means that our findings cannot be generalized to all institutions or sectors in Bangladesh. For example, our analysis focused primarily on urban, research-oriented universities with emerging AI programs, which may obscure practices in smaller regional or vocational institutions where different challenges and resources shape readiness. Similarly, although we sought to balance voices across faculty, students, and industry professionals, some groups—particularly policymakers—were less accessible, limiting our ability to fully triangulate government perspectives. Moreover, our qualitative methods emphasize depth of experience and contextual analysis over statistical breadth; as a result, the insights we provide should be understood as analytic generalizations rather than predictive claims.

Second, our findings are also shaped by the temporal and political context of data collection. AI education in Bangladesh is evolving rapidly, with new policy initiatives, international partnerships, and industry investments underway; some of our observations may shift as these infrastructures change. Our reliance on interviews and self-reported experiences also introduces the possibility of social desirability bias, particularly when participants reflected on their own institutions. Additionally, our application of STS concepts such as blackboxing, strain, and path dependency necessarily involves interpretive framing; while these lenses enabled rich theorization, they may also highlight certain dynamics while downplaying others. Finally, our study foregrounds Bangladesh as a case site, and while many insights may resonate with other Global South contexts, local specificities—such as the role of the University Grants Commission or the cultural politics of gendered mentorship—may not translate directly. These limitations do not undermine the contributions of the study but rather indicate the need for continued, comparative research across institutions, time, and geographies.

\section{Future Work}
Building on these limitations, future work should pursue three directions. First, there is a need for \textbf{comparative studies across Global South contexts}. While our findings in Bangladesh reveal how infrastructural scarcity, blackboxing, strain, and path dependency shape AI education, similar dynamics may manifest differently in countries such as India, Nigeria, or Brazil. Comparative research can help identify which challenges are globally recurrent and which are locally specific, offering insights into how sociotechnical infrastructures of readiness are configured across diverse contexts.

Second, future research should adopt \textbf{longitudinal methods} to capture how AI readiness evolves over time. Curricular reforms, policy initiatives, and new industry partnerships are dynamic processes that may reshape readiness in ways not visible in a single snapshot. Ethnographic follow-ups, repeated surveys, or institutional case histories would allow scholars to trace how infrastructural interventions succeed, stall, or transform, thereby expanding our understanding of path dependency and repair.

Third, there is an urgent need for \textbf{intervention-driven research}. Our study identifies actionable design solutions—such as de-blackboxing mentorship, modular ethics toolkits, and cooperative GPU labs—but testing these interventions in practice remains a critical next step. Future HCI and ICTD research could engage in participatory co-design with faculty, students, and policymakers to prototype and evaluate these solutions. Such interventions would not only generate practical improvements but also refine theory, revealing how infrastructural, social, and postcolonial dynamics can be shifted through intentional design.

By pursuing comparative, longitudinal, and intervention-driven research, the field can deepen its understanding of AI readiness as a sociotechnical and postcolonial condition. This future work will help ensure that readiness is not framed narrowly as ``catching up'' with Euro-American benchmarks but instead as cultivating infrastructures of equity, responsibility, and locally grounded innovation.

\begin{acks}
Anonymized.
\end{acks}



\bibliographystyle{ACM-Reference-Format}
\bibliography{sample-base}

\end{document}